\documentclass[a4paper,11pt]{extarticle}
\usepackage{geometry}
\usepackage[T1]{fontenc} 
\usepackage[utf8]{inputenc}
\usepackage{authblk}

\usepackage{color}
\usepackage[numbers,sort&compress]{natbib}
\usepackage[colorlinks,citecolor=blue,urlcolor=blue,linkcolor=blue]{hyperref}
\usepackage{graphicx}
\usepackage{mathptmx}      
\usepackage{flushend}

\usepackage{amssymb}
\usepackage{amsmath}
\usepackage{placeins}
\usepackage{bm}
\usepackage{hyperref}
\usepackage{mathtools}
\usepackage{xcolor}
\usepackage{soul}

\usepackage{multirow}
\usepackage{lipsum}
\usepackage{makecell}
\usepackage{adjustbox}

\usepackage{fancyhdr}
\fancypagestyle{plain}{%
	\fancyhf{}
	\rhead{\thepage}
}

\begin{document}

        \title{\bf Dynamical Wormhole Solutions in Rastall Theory}
        
        \author[1]{Yaghoub Heydarzade\thanks{\href{mailto:yheydarzade@bilkent.edu.tr}{yheydarzade@bilkent.edu.tr}}}
        \author[2]{Maryam Ranjbar\thanks{\href{mailto:maryamrnjbr96@gmail.com}{maryamrnjbr96@gmail.com}}}
        
        \affil[1]{Department of Mathematics, Faculty of Sciences, Bilkent University, 06800 Ankara, Turkey}
        \affil[2]{Istanbul, Turkey}
        
        \date{\today}

\maketitle

\begin{abstract}
        Wormhole configurations in Einstein's general theory of relativity (GR) require exotic matter sources violating the weak energy condition (WEC). Rastall's theory is a generalization of GR in its matter source considering a nonconserved energy-momentum (EM) tensor. Hence, on one hand, the nature of this generalization of the matter source of field equations and, on the other hand, the possibility of respecting energy conditions for dynamical wormholes in contrast to static ones motivates us to study the possibility of the existence of wormhole configurations respecting energy conditions or minimizing the violations of them in Rastall's modified theory.   We derive general analytical solutions considering a constant redshift function and a particular equation of state for energy density and pressure profiles. We show that because of the modification in the EM source of the field equations, there exist solutions respecting the WEC in the vicinity of the wormhole's
        throat for specified values of the parameters. Some particular solutions are discussed in detail.
\end{abstract}

\maketitle

\section{\label{sec:Intro}Introduction\protect}
Despite the success of Einstein's general relativity (GR) in explaining many gravitational phenomena, it falls short in explaining dark matter and dark energy. To address these issues, modifications of GR have been proposed, e.g scalar-tensor theories \cite{intro25}, f(R) theories \cite{intro29}, and braneworlds \cite{intro31}. For a comprehensive review, see  \cite{intro24}.

 In 1972, Peter Rastall proposed a modification to Einstein's theory with a nonconserved energy-momentum
tensor \cite{intro36}. In his theory, the divergence of energy-momentum tensor is proportional to the gradient of the Ricci scalar through a proportionality constant \cite{intro36}. Hence,  in contrast to the standard conservation law of energy-momentum, the Bianchi identity still holds. Rastall gravity yields some interesting results, for instance, the late time accelerating expansion of the universe can be explained \cite{intro37}, and the de Sitter black hole solutions can be found without explicitly assuming a cosmological constant \cite{introbh}.  
The question of equivalence of Rastall gravity to Einstein's theory as a redefinition of the EM tensor was raised in  \cite{introvis}. However,
it has been shown that the nature of this theory considering a nonconserved
EM source is not just a redefinition of EM,  and it gives different results
than GR, see for instances \cite{intro38, intro371, intro372, intro373, intro374}.
It is shown recently that a Lagrangian formulation for a Rastall-type theory can be provided in the context of $f(R, \mathcal{L}_m)$ and $f(R, T)$ theory \cite{introlag, intro381} where $R$ is the Ricci scalar, $\mathcal{L}_m$ is the Lagrangian of matter fields and $T$ is the trace of the energy-momentum tensor.

Einstein's general theory of relativity (GR) admits solutions describing geometrical bridges connecting two distant regions of a universe or even two different universes. For the first time it was Wheeler who proposed the term "wormhole" for these geometrical bridges  in order to provide a mechanism for having "charge without charge". He claimed that the electric charge emerges as a manifestation
of the topology of a space, a sheet with a handle \cite{introwheel}.  The interest in these solutions almost declined over years until the notion of traversable Lorentzian wormholes was introduced by Morris, Thorne and Yurtsewer \cite{introMTY1, introMTY2}. It was discussed that these structures could allow humans not only to travel between distant parts of a universes, or even two universes, but also to construct time machines.
In the framework of GR, the {\it flaring-out condition} on the throat of wormhole leads to the violation of weak energy condition (WEC) demanding an {\it exotic matter} source in the Earth-based laboratory context. This violation of the energy condition is conventionally a problematic issue that requires a resolution or at least a minimization \cite{WHbook, Hochberg-Visser, Hochberg-Visser2, intro64}. 
Numerous studies have endeavored to address the nature of exotic matter within various settings \cite{viser-d-1989, viser-n-1989, Eiroa-2005, Zaslavski-2007}. One approach is to
construct thin-shell wormholes in the context of GR via cut-and-paste procedure in which the exotic matter source is minimized by concentrating at the wormhole's throat \cite{viser-d-1989, introEC2, introor1, introor2}.
Another approach is to investigate the modified
theories of gravity where the presence of curvature higher order terms in curvature may provide a possibility for constructing wormhole structures by ordinary matter sources \cite{ lobo-2011, harko-2013}. As instances, see wormhole solutions in Brans-Dicke theory \cite{intro52, intro53, intro54, intro55, bran}, Einstein-Gauss-Bonnet theory
\cite{dotti-2007, dotti-2009}, $f(R)$ gravity \cite{intro56, intro57, intro58, Bhattacharya-2017}, and scalar-tensor gravity \cite{intro59}, higher dimensioanal
theories \cite{camera-2003, higherdim-dotti, Lovelock-gravity, Torii-2013,
dgp}. Moreover, in contrast to static wormholes in GR, it has been noted that for evolving wormholes there is this possibility of satisfaction of energy conditions for a
finite interval of time \cite{intro62, intro63}, see also pioneer works \cite{intro64,intro65, intro66, intro67, intro68,intro69,intro71,intro73,intro74,intro75,intro76,intro79}.
\\
Akin to the other modified theories,  Rastall theory also has numerous successful applications in cosmology and astrophysics and this drives a motivation for investigating it versus the conditions for the existence wormholes structures. Nevertheless, our main motivation for the present study rely on very distinct feature of this theory that distinguishes it from other modified
theories:  its modification in the matter source of the Einstein field equations only and leaving the geometric part unaltered. As the result, this provides very unique possibilities in
the context of this theory as i) the field equations
remain rather simple to be handled, and ii) the main concern in constructing
wormholes, need for exotic matter fields, can be traced easily by the nonminimal coupling of the EM tensor and geometry, and their interplay through a constant
coupling  parameter. We will see the footprint of this coupling in the solutions derived. On the other hand, the possibility of respecting ECs for finite
time intervals in dynamical configurations, in contrast to static cases, in GR, stimulate another motivation to investigate how these dynamical configurations
behave in Rastall theory. Therefore, the objective of our study is to discover viable  dynamical wormhole solutions within the framework of Rastall theory and demonstrate how the nonminimal coupling nature of this theory influences the shape and evolution of these solutions. Here it is necessary to to mention that static wormhole solutions have been studied in the context of Rastall theory showing that the WEC can be met for some particular solutions, see for instance  \cite{introras2, introras3, lob, naz, traversable}. In \cite{introras4} it is shown that Rastall theory is capable of modifying the energy condition requirements of the matter source to satisfy the strong energy condition at the throat. This modification demands that either the Rastall coupling $ \kappa$ or $\lambda$ has to be negative.  It is concluded that Rastall gravity has the potential to alleviate some issues encountered by static wormholes within the framework of Einstein gravity. 
Since the dynamical wormholes in the context of Rastall theory have not been studied yet, it seems worthwhile to put one step further to explore the theory for the possible generalizations of the static solutions to dynamical cases.

 The organization of the paper is as follows. In section II, we derive the general analytical solutions of the field equations for a wormhole geometry. In section III, we analyse some particular solutions versus the flaring out and  WEC, and show that under some constraints these conditions are respected in the context of Rastall gravity. Section IV is devoted to our concluding remarks.

\section{\label{sec:WH1}Evolving wormholes in Rastall Theory\protect}
The validity of the energy-momentum conservation law in the four dimensional spacetime was questioned by Rastall \cite{intro36}. He considered the following hypothesis
\begin{equation}
        \label{13}
        \mathit{T}^{\mu \nu}\, _{; \mu}= \lambda \mathfrak{R}^{, \nu},
\end{equation}
where $ T^{\mu\nu}$
is the energy-momentum tensor of matter source,  $\lambda$  is the Rastall constant parameter, and $\mathfrak{R}$ is the Ricci scalar. Hence, the Einstein field equations get modified as
\begin{equation}
        \label{14}
        \mathit{G}_{\mu \nu}+ \kappa \lambda g_{\mu \nu} \mathfrak{R} = \kappa \mathit{T}_{\mu \nu}, 
\end{equation}
where $\kappa$ is the gravitational coupling. In the present work, we are interested in dynamical wormhole solutions of these field equations. For the static wormhole solutions in Rastall theory, see  \cite{introras2, lob, naz, introras3, introras4, traversable}. Hence, we consider time-dependent generalization of Morris-Thorne wormhole metric as \cite{introMTY1} 
\begin{equation}
        \label{1}
        ds^2 = -U(r) dt^2 + \mathit{R}(t)^2 \left( \frac{dr^2}{1-\frac{B(r)}{r}} + r^2 (d \theta^2 + sin^2 \theta \, d\phi^2) \right),
\end{equation}
where $R(t)$ is the scale factor of the background Universe, $U(r)$ is the redshift function and $B(r)$ is the wormhole shape function.  The static Morris-Thorne wormhole is recovered by setting $R(t)=constant$. In order to have a wormhole geometry,  the following general constraints on the redshift and shape functions are required \cite{introMTY1, introMTY2}.
\begin{itemize}
        \item[$\bullet$] The wormhole throat connecting two asymptotic regions is located at the minimum radial coordinate  $r_0=B(r_0)$.
        \item[$\bullet$] The shape function $B(r)$ must satisfy the so-called flaring-out condition  $B(r)-rB^\prime (r)>0$ at the vicinity of the throat which reduces to $B^{'}(r_0)<1$ at the throat.
        \item[$\bullet$]  In order to keep the signature of the metric for $r > r_0$,  the shape function holds the condition $1-\frac{B(r)}{r}>0$.
        \item[$\bullet$] For asymptotically flat
        wormholes, the metric functions should satisfy the conditions $U(r)\to 1$, ~ $B(r)/r \to 0$ as $r \to \infty$. In this case,  the metric (\ref{1}) tends
        to the flat Friedmann-Robertson-Walker metric in the asymptotic region.
        
        \item[$\bullet$] The redshift function $U(r)$ must be finite and nonzero throughout the spacetime in order to ensure the absence of horizons and singularities. 
        
\end{itemize}
We use a similar methodology as in  \cite{Bhat-2021}
for evolving Lorentzian wormholes in GR. We will see that how Rastall's paprameter appears in the solutions for the scale factor and shape function to modify the similar solutions in \cite{Bhat-2021}. Considering the metric ({\ref{1}}) with the constant redshift function $U(r)=1$, and the energy-momentum tensor $T^{\mu}_{\nu}=diag\left(-\rho(t,r), P_r(t,r), P_l(t,r), P_l(t,r)\right)$, field equations (\ref{14}) yield
\begin{equation}
        \label{17}
        \rho(t,r)=\frac{1}{\kappa} \Big( 3 \mathrm{H}^2 + \frac{B^{'}(r)}{r^2 \mathit{R}(t)^2} -\kappa \lambda \mathfrak{R} \Big),
\end{equation}
\begin{equation}
        \label{18}
        P_r(t,r)= \frac{1}{\kappa} \Big( -3 H^2 -2 \dot{H} -\frac{B(r)}{\mathit{R}(t)^2 r^3} + \kappa \lambda \mathfrak{R} \Big),
\end{equation}
\begin{equation}
        \label{19}
        P_l(t,r)= \frac{1}{\kappa} \Big( -3H^2 - 2\dot{H} -\frac{B^{'}(r)}{2 r^2 \mathit{R}(t)^2} +\frac{B(r)}{2 r^3 \mathit{R}(t)^2} + \kappa \lambda \mathfrak{R} \Big),
\end{equation}
where $H=\dot R(t)/R(t)$, and the Ricci scalar reads as 
\begin{equation}
        \label{20}
        \mathfrak{R}=\frac{2 B^{'}(r)}{r^2 \mathit{R}(t)^2} + 12 \mathit{H}^2 + 6 \dot{\mathit{H}}.
\end{equation}
For integrating the present system of three nonlinear partial differential equations (\ref{17}), (\ref{18}) and (\ref{19}) with five unknowns $R(t), B(r), \rho(t,r), P_r(t,r) $ and
$P_l(t,r)$, one can consider a physically motivated constraint; more specifically
an equation of state for the sets of unknowns $(\rho(t,r), P_r(t,r))$ and $(\rho(t,r), P_l(t,r))$ or even for $(P_r(t,r), P_l(t,r))$ as in \cite{Bhat-2021}. Another possibility is to consider traceless
constraint on EM tensor as in \cite{kar}.  Here, in order to keep the equation
of state as much as possible general which can reduce to some known specific equations
of state, we consider a general EoS including our three unknowns $(\rho(t,r), P_r(t,r),P_l(t,r)$ as in \cite{EOS1, EOS2}
\begin{equation}
        \label{11}
        \rho(t,r)=\frac{\omega}{1+2 \gamma} \Big(P_r (t,r)+ 2 \gamma P_l(t,r) \Big),
\end{equation}
where $\omega$ and $\gamma$ are equation of state parameters. This equation
of state depending on two parameters $\omega$ and $\gamma$ can  reduce to
the following special cases: $i)$ the barotropic EoS as $\rho(t,r)=\omega P(t,r)$ when $P_r(t,r)=P_l(t,r)=P(t,r)$, $\forall \gamma$, which reduces
to cosmological constant for $\omega=-1$, ~~$ ii)$ the traceless EM's  EoS as $-\rho(t,r) +P_r(t,r)+2P_l(t,r)=0$ when $\omega=3, \, \gamma=1$,~~ and $iii)$ the dimension $(n)$ dependant EoS $\rho(t,r)=\alpha\left(P_r (t,r)+ (n-2) P_l(t,r) \right)$  \cite{dim} in $n=4$ when $\gamma=1$. Later we will see that how
the Rastall's coupling $\beta$  and the wormhole conditions  together put
constraints on each of these two parameters $\omega$ and $\gamma$ in \eqref{11}.

Combining the set of equations (\ref{17}, \ref{18}, \ref{19}) with the EoS (\ref{11}), we obtain the following single nonlinear partial differential equation in our unknown functions $B(r)$ and $R(t)$
\begin{eqnarray}
        \label{22}
                \frac{ \left( 1+ \gamma (2 + \omega) \right) r B^{'}(r) \, - \omega (\gamma -1) B(r)}{ \kappa (1+2 \gamma)r^3} &=&  - \frac{\mathit{R}(t)^2 \, (1+2 \gamma) \left( 8 \omega \dot{\mathit{H}} + 12 H^2 (\omega +1) \right)}{(4+ 8 \gamma) \kappa}\nonumber\\
                &&+ \lambda \mathfrak{R} (1+ \omega) \mathit{R}(t)^2.
\end{eqnarray}
This equation can be integrated for $B(r)$ and $R(t)$ by separating it into the radial and temporal parts  as follows
\begin{equation}
        \label{23}
        \begin{split}
                &~ \frac{ \left( 1+ \gamma (2 + \omega) \right) r B^{'}(r) \, - \omega (\gamma -1) B(r)}{ (1+2 \gamma)r^3}  -\frac{2\beta (1+\omega) B^{'}(r)}{r^2}=  \\
                &~~\beta (1+ \omega) \mathit{R}(t)^2 (12 \mathit{H}^2 + 6 \dot{\mathit{H})} - \frac{\mathit{R}(t)^2 \, (1+2 \gamma) \left( 8 \omega \dot{\mathit{H}} + 12 H^2 (\omega +1) \right)}{(4+ 8 \gamma) },
        \end{split}
\end{equation}
where $\beta=\kappa \lambda$. This equation can be considered as the master equation to be solved for our unknowns, and
it is similar to the master equation in  \cite{Bhat-2021}. In \cite{Bhat-2021} the master equation was
derived by combining the field equations considering the relation  $p_r(t,r)=\alpha p_t(r,t)$ where in general $\alpha=\alpha(r)$. However, one notes to the modification here by the Rastall's parameter $\beta$ and the difference in the coefficients due to the different equation of states used.  The radial and temporal parts of Eq. (\ref{23}) give the following ordinary differential
equations (ODEs) for the shape function and scale factor respectively
\begin{equation}
        \label{270}
        \frac{ \Big( 1+ \gamma (2 + \omega) \Big) r B^{'}(r) \, - \omega (\gamma -1) B(r)}{ (1+2 \gamma)r^3}  -\frac{2\beta(1+\omega) B^{'}(r)}{r^2} =C,
\end{equation}
and 
\begin{equation}
        \label{27}
        R(t)^2 \left[\left(6 \beta (\omega +1) - 2 \omega \right) \dot{H}+\left(12 \beta (\omega +1)- 3(\omega +1) \right)H^2 \right] =C.
\end{equation}
Let the constants $a=6 \beta (\omega+1)-2 \omega$ and $d= 12 \beta (\omega+1)-3(1+\omega)$, then Eq.(\ref{27}) can be rewritten as
\begin{equation}
        \label{28}
        R(t)^2 [a \dot{H}+ d H^2]= C,
\end{equation}
or equivalently 
\begin{equation}
        \label{31}
        a R(t) \ddot{R}(t)+ b \dot{R}(t)^2 =C,
\end{equation}
where the constant $b= d - a = a + \omega -3$. Here, one notes that the dynamics of the scale factor depends on the Rastall's coupling parameter $\beta$ and EoS parameter $\omega$ while is independent of the parameter $\gamma$. \\
In the following subsections, we obtain general exact solutions to Eqs.(\ref{270}) and (\ref{31}) for two cases $C=0$ and $C\ne 0$. Some particular sub-classes of the obtained general solutions will be investigated versus the flaring out and weak energy conditions in the next section.
\subsection{\label{sol-c1=0} Solutions for  $C=0$}
\subsubsection{\label{shapefunction0}Solution for the shape function}
Integrating Eq.(\ref{270}) for $C=0$, the shape function can be obtained as
\begin{equation}
        \label{32}
        B(r)= r_0 \left(\frac{r_0}{r}\right)^{\frac{(1-\gamma ) \omega}{1-2 \beta (2 \gamma +1)  (\omega+1)+\gamma ( \omega+2)}},
\end{equation}
Here one observes that how the Rastall's coupling parameter $\beta$ modifies the wormhole's shape function in comparison to the case of GR when $\beta=0$. The resulting geometry can be asymptotically flat  or nonflat depending on the set of
parameters $\omega,\gamma$ and $\beta$. 

The flaring out condition at the throat reads as
\begin{equation}
        \label{33}
        B'(r_0)=\frac{(\gamma -1) \omega}{1-2 \beta (2 \gamma +1)  (\omega+1)+\gamma ( \omega+2)}<1.
\end{equation}
Moreover,  in order to satisfy the asymptotically flatness  $\frac{B(r)}{r}\to 0$ as $r\to \infty$, the following condition should be fulfilled
\begin{equation}
        \label{flat1}
        -1<\frac{(1-\gamma ) \omega}{1-2 \beta (2 \gamma +1)  (\omega+1)+\gamma ( \omega+2)}<1.
\end{equation}

\subsubsection{\label{scalefactorc1=0}Solution for the scale factor}
One can integrate Eq.(\ref{31}) for $C= 0$ to find the general solution 
\begin{equation}
        \label{57.1}
        R(t)= \left(R_0 \, t+ R_1\right)^{\frac{1}{1+ b/a}}=\left(R_0 \, t+ R_1\right)^{\frac{a}{d}},
\end{equation}
where $R_0$ and $R_1$ are integration constants. One observes that this solution does not contain the Big Bang singularity if $t\neq -\frac{R_1}{R_0}$. Here
one notes  that the solution \eqref{57.1} is a generic dynamic wormhole solution that is similar to the solution obtained in  \cite{Bhat-2021} in GR. Hence,
the general form of the solution for the scale factor is independent of the Rastall gravity due to the similarity in the governing ODE on $R(t)$ in \eqref{31}. However the solutions may differ depending on the assumed parameter constraints for the purpose of the solution in the underlying theory. Here, Rastall's coupling $\beta$ arises in the power $\frac{a}{d}$ and can be considered as a factor for distinguishing the solution from those in GR in the limit $\beta\to 0$. Later we will discuss  the values of $\beta$  parameter and its effect in satisfaction of wormhole conditions.
The following particular subclasses of (\ref{32}) and (\ref{57.1})  can be of interest.
\begin{itemize}
        \item[$\bullet$] \label{r1} \textit{ $\mathbf{a=d}$}\\
        \\
        For this case, the scale factor, shape function and $\omega$ are given  by
        \begin{equation}
                \label{61.1}
                \begin{split}
                        & R(t)=R_0 \,t +R_1,~~~~ \omega =\frac{6 \beta -3}{1-6 \beta },~~~~ \beta \ne \frac{1}{6},\\
                        & B(r)=\frac{r^3}{r_0^2}.
                \end{split}
        \end{equation}
        One can verify that this solution to  (\ref{23})  fails to satisfy the flaring out condition for evolving wormhole solutions. Hence, we do not analyze this solution versus the WEC.
        \item[$\bullet$] \label{r2} 
        \textit{$\mathbf{a=2d}$}\\
        \\
        In this case, we have
        \begin{equation}
                \label{67.1}
                \begin{split}
                        &R(t)=\left(R_0 t + R_1\right)^2,~~~~\omega =\frac{9 \beta -3}{2-9 \beta },~~~~ \beta \ne \frac{2}{9},\\
                        & B(r)=r_0 \left(\frac{r_0}{r}\right)^{\frac{3 (3 \beta -1) (\gamma -1)}{\beta  (5 \gamma +7)-\gamma -2}},
                \end{split}
        \end{equation}
        where $\gamma$ should satisfy the wormhole conditions.
        \\
        \item[$\bullet$] \label{r3} 
        \textit{$\mathbf{a=\frac{1}{2} d}$}\\
        \\
        In this case, we find
        \begin{equation}
                \label{73.1}
                \begin{split}
                        & R(t)=\left(R_0 t + R_1\right)^{\frac{1}{2}},~~~~\omega=3,\\
                        & B(r)=r_0 \left(\frac{r_0}{r}\right)^{\frac{3 (\gamma -1)}{8 \beta  (2 \gamma +1)-5 \gamma -1}}.
                \end{split}
        \end{equation}
        Here, $\beta$ parameter remains arbitrary and  $\gamma$ should satisfy the wormhole conditions. Here to make clear how the Rastall gravity, and not only the choice of the stress-tensor, is important in influencing the solutions \eqref{67.1} and \eqref{73.1}, one may consider the following two possibilities: $i)$ fix the  parameter $\gamma$ and $\omega$  by assuming known specific stress energy tensors at this step, so that solutions now clearly depend on the Rastall factor, and $ii)$ consider the theoretically and observationally verified values or ranges on
Rastall parameter $\beta$, and then obtain corresponding allowable $\omega$ and $\gamma$ values satisfying the wormhole conditions that can include parameter ranges for both the normal and exotic matters. The latter possibility implies
how the coupling parameter $\beta$ confines or affects the matter sources needed for such configurations. Up to this  point, one observes the constraint on $\omega$ parameter. In section 3, in order to investigate the obtained viable solutions versus the ECs, regarding the theoretical and observational constraints on $\beta$ parameter  \cite{intro36, PRD-moradpour,El-Hanafy, stz967, moradpur-beta1.6}, we will consider two admissible ranges $0<\beta<\frac{1}{6}$ and $\beta<0$, and we  will analyse the above latter possibility in detail. Specifically,
we show that the satisfaction of wormhole conditions is possible for two
observationally obtained values of $\beta=0.163$ \cite{stz967} and $\beta=0.041$
\cite{ El-Hanafy}. As an instance, for the particular solution of
$a=\frac{1}{2}d$, consideration the EoS parameters $\omega=3, \gamma=0.35$ with $\beta=0.041$ provides the possibility of satisfaction of all wormhole conditions that is illustrated in Figure \ref{fig.6}. This is an interesting case in the sense that substituting these EoS parameters in \eqref{11} and defining an effective pressure $P_e(t,r)=P_r(t,r)+(0.7) P_l(t,r)$ we have an effective equation of state $P_e(t,r)=\frac{1.7}{3}\rho(t,r)$  which denotes a matter source respecting ECs. This indeed is an example for the first possibility mentioned
above as well.
\end{itemize}

\subsection{\label{sol-c1ne0} Solutions for  $C\ne 0$}
\subsubsection{\label{shapefunctioncne0} Solution for the shape function}
The shape function $B(r)$ can be obtained by integrating (\ref{270}) as
\begin{equation}
        \label{24}
        B(r)= -\frac{C }{6 \beta  (\omega+1)-\omega-3}r^3+ C_1 \, r^{\frac{(\gamma -1) \omega}{1-2 \beta (2 \gamma +1) (\omega+1)+\gamma ( \omega+2)}},
\end{equation}
where $C$ and $C_1$ are separation and integration constants, respectively. Like \eqref{57.1}, the solution
 \eqref{24} is a generic wormhole shape function and is similar to the solution in  \cite{Bhat-2021}. The difference being is upto some parameter choices. However, one observes that, as we will see later in analyzing solutions versus
WEC, the difference in the underlying theories, i.e here the being of Rastall parameter $\beta$, can play a crucial role in satisfying wormhole conditions even by ordinary matter sources. This indeed implies how such a modification in EM source, akin to the higher order curvature terms in other modified theories,  is capable of solving the issue of the need for exotic matter in GR.
 To be specific, the presence of $\beta$ puts constraints on the required matter sources, i.e on $\omega$ and $\gamma$, see the classification given
in Table \ref{tab1}. In other words, as discussed in \cite{introras4} for
static cases, considering the field equations $G_{\mu\nu}=\kappa_r S_{\mu\nu}$ where
the effective EM tensor $S_{\mu\nu}$ includes the Rastall's modification term $\beta \mathfrak{R}g_{\mu \nu}  $,  the actual matters make up with phantom characteristics. Therefore, in Rastall gravity, general wormhole solutions can exist with both normal and phantom matter, depending on the Rastall coupling parameter. 

Using the (initial) condition
$B(r_0)=r_0$ at the wormhole's throat we can determine integration constant $C_1$ as
\begin{equation}
        \label{25}
        C_1= \frac{\left(6 \beta (\omega+1)-\omega-3\right)r_0 + C  r_0^3}{\left(6 \beta  (\omega+1)- \omega - 3\right)r_0^{\frac{(\gamma -1) \omega}{1-2 \beta (2 \gamma +1)  (\omega+1)+\gamma ( \omega+2)}}},
\end{equation}
from which we find the flaring out condition at the throat
as\begin{equation}
        \label{26}
        B'(r_0)=\frac{ -C r_0^2(1+2 \gamma)  +\omega(1-\gamma) }{-1+2 \beta (2 \gamma +1)   (\omega+1)-\gamma ( \omega+2)}<1.
\end{equation}
Here one observes that depending on the set of parameters $ \omega,\gamma$, and $\beta$, the coefficient of the first term in \eqref{24},
i.e $k=\frac{C }{6 \beta  (\omega+1)-\omega-3}$, appears as an effective cosmological constant. This  means that for $C\neq 0$, we have asymptotically  (anti) de Sitter-like solutions and  the asymptotic flatness condition
does not hold here. Also, as it is pointed out in  \cite{Bhat-2021}, the
above defined $k$ constant can be interpreted as a topological
number denoting the spatial curvature of the background FRW spacetime taking
values $\pm 1, 0$ representing a closed, open and flat universe, respectively.  one can
write the $B(r)$ function as
\begin{equation}
        \label{24-2}
        B(r)= -k r^3 + B_n(r),
\end{equation}
 where $k$ represents the spatial curvature of the FRW metric
 and  $B_n(r)$ is the shape function of a wormhole inhabiting within this spacetime. One should  note to the difference here in \eqref{25} and \eqref{26}, similar to \cite{ext1, ext2} as instances, and in \cite{Bhat-2021} where the throat condition $B_n(r_0)=r_0$  is imposed only on the second term $B_n(r)$ in the shape function. It is mentioned in \cite{ext1, ext2}
that imposing the throat condition $B(r_0)=r_0$, the spatial extension of the wormhole solution cannot be arbitrarily large.
 Following \cite{Bhat-2021}, the throat condition $B_n(r_0)=r_0$ together with
 t the flaring out condition
give
\begin{equation}
        \label{33-2}
        B'(r_0)=\frac{(\gamma -1) \omega}{1-2 \beta (2 \gamma +1)  (\omega+1)+\gamma ( \omega+2)}<1.
\end{equation}
The asymptotic flatness condition reads as 
\begin{equation}
        \label{flat1-2}
        -1<\frac{(1-\gamma ) \omega}{1-2 \beta (2 \gamma +1)  (\omega+1)+\gamma ( \omega+2)}<1.
\end{equation}

\subsubsection{\label{scalefactorc1ne0} Solution for the scale factor}
Considering the general case $a,b\ne 0$, Eq.(\ref{31}) can be integrated giving the following first order nonlinear differential equation
\begin{equation}
        \label{98}
        \dot{R}^2(t)=\frac{C}{b} \left( 1-R_0 R^{- \frac{2b}{a}}\right),
\end{equation}
where $R_0$ is an integration constant, and hence
\begin{equation}
        \label{99}
        \int \frac{d R}{\sqrt{1-R_0 R^{ -\frac{2b}{a}}}}= \pm \sqrt{\frac{C}{b}}\int d t,
\end{equation}
for $C/b>0$. Here one can obtain the explicit from  of the scale factor $R(t)$ for some particular cases of parameters $a$ and $b$. 
The following particular cases can be of interest.
\begin{itemize}
        \item[$\bullet$]  \textit{ $\mathbf{a=-2b}$}\\
        
        This case gives the scale factor $R(t)$, $\omega$ and shape function $B(r)$ as follows
        \begin{equation}
                \label{101}
                \begin{split}
                        R(t)& =\frac{1}{R_0}-\frac{R_0}{4}\left(\pm \sqrt{\frac{C}{b}}
                        t + R_1\right)^2,~\omega =\frac{3-9 \beta }{9 \beta -2},~~\beta \ne \frac{2}{9}, \\ 
                        B(r) & =\frac{(9 \beta -2) C }{12 \beta -3}r^3 +\frac{r_0 \left((3-12 \beta )+(9 \beta -2) C r_0^2\right)}{12 \beta -3} \, \left(\frac{r}{r_0}\right)^{\frac{3 (3 \beta -1) (\gamma -1)}{-\beta  (5 \gamma +7)+\gamma +2}},
                \end{split}
        \end{equation}
        where $R_1$ is an integration constant.\\
        Considering $B_n(r)$ as the shape function of the
        inhabiting wormhole, we have
        \begin{equation}
                \label{101-2}
                \begin{split}
                        R(t)& =\frac{1}{R_0}-\frac{R_0}{4}\left(\pm \sqrt{k}
                        t + R_1\right)^2,~\omega =\frac{3-9 \beta }{9 \beta -2},~~\beta \ne \frac{2}{9}, \\
                        B_n(r) & = r_0
                        \, \left(\frac{r}{r_0}\right)^{\frac{3 (3 \beta -1) (\gamma -1)}{-\beta  (5 \gamma +7)+\gamma +2}},
                \end{split}
    \end{equation} 
where the reality of the solution requires $k=1$. Later we will show that the WEC can be respected in both the above cases for $a=-2b$. \item[$\bullet$] \textit{ $\mathbf{a=-b}$}\\In this case, one finds
        \begin{equation}
                \label{102}
                \begin{split}
                        & R(t)= \frac{1}{\sqrt{R_0}}\sin \left(\pm \sqrt{\frac{CR_0}{b}}\,t+R_1\right),~~~\beta = \frac{1}{4},\\
                        & B(r)= -\frac{2 C }{\omega -3}r^3 + \frac{r_0 \left(2 C r_0^2+\omega -3\right)}{\omega -3} \, \left(\frac{r}{r_0}\right)^{\frac{2 (\gamma -1) \omega }{2 \gamma -\omega +1}},
                \end{split}
        \end{equation}
        where $R_1$ is an integration constant. We do not analyze this solution versus the wormhole conditions since the contraction of the field equations  (\ref{14}) by the metric gives the Ricci scalar as $\mathfrak{R}=\frac{1}{1-4\beta}T$ which diverges for $\beta = \frac{1}{4}$ and $T\neq  0$ \cite{intro36}.
\end{itemize}
\section{\label{sec:WEC} Weak Energy Condition}
In order to investigate the obtained viable solutions versus the energy conditions, regarding the theoretical and observational constraints on $\beta$ parameter  \cite{intro36, PRD-moradpour,El-Hanafy, stz967, moradpur-beta1.6}, we will consider two admissible ranges $0<\beta<\frac{1}{6}$ and $\beta<0$.
\subsection{\label{beta>0} WEC for $0<\beta<\frac{1}{6}$}
In this subsection, considering $0<\beta<\frac{1}{6}$ we obtain the valid ranges of $\omega$ and $\gamma$ satisfying both the WEC ($\rho \ge 0$, $\rho+P_r >0$ and $\rho+P_l>0$) and flaring-out condition ($B^{'}(r_0)<1$) simultaneously.
\subsubsection{\label{wec>0-c=0} Analysis of solutions for $C=0$}
Here we analyze the following particular solutions for the scale factor when $C=0$.

\begin{itemize}
        \item[$\bullet$] \label{rr2} 
        \textit{$\boldsymbol{a}=\boldsymbol{2 d}$}\\
        
      Inserting the scale factor and the shape function
      in (\ref{67.1})  into the field equations (\ref{17}-\ref{19}), one obtains
        
        \begin{equation}
                \begin{split}
                        \label{69.1}
                        \rho(t,r)=& -\frac{3 (3 \beta -1) (6 \beta -1) R_0^2}{2 \pi G (4 \beta -1) (R_0 t+R_1)^2} \\
                        & + \frac{3 r_0^{-2} (2 \beta -1) (3 \beta -1) (6 \beta -1) (\gamma -1)}{8 \pi G (4 \beta -1)  (\beta  (5 \gamma +7)-\gamma -2) (R_0 t+R_1)^4} \, \left(\frac{r_0}{r}\right)^{3+ \frac{3 (3 \beta -1) (\gamma -1)}{\beta  (5 \gamma +7)-\gamma -2}},
                \end{split}
        \end{equation}
        
        \begin{equation}
                \begin{split}
                        \label{70.1}
                        \rho(t,r)+P_r(t,r)& =\frac{(6 \beta -1) R_0^2}{2 \pi G (4 \beta -1)(R_0 t+R_1)^2}\\
                        & -\frac{  r_0^{-2} (6 \beta -1) (2 \beta  (7 \gamma -1)-4 \gamma +1)}{8 \pi G (4 \beta -1) (\beta  (5 \gamma +7)-\gamma -2) (R_0 t+R_1)^4} \, \left(\frac{r_0}{r}\right)^{3+ \frac{3 (3 \beta -1) (\gamma -1)}{\beta  (5 \gamma +7)-\gamma -2}},
                \end{split}
        \end{equation}
        
        \begin{equation}
                \begin{split}
                        \label{71.1}
                        \rho(t,r)+P_l(t,r)& =\frac{(6 \beta -1) R_0^2}{2 \pi G (4 \beta -1)(R_0 t+R_1)^2}\\
                        &-\frac{  r_0^{-2} (6 \beta -1) (4 \beta  (\gamma -4)-2 \gamma +5)}{16 \pi G (4 \beta -1) (\beta  (5 \gamma +7)-\gamma -2) (R_0 t+R_1)^4} \, \left(\frac{r_0}{r}\right)^{3+ \frac{3 (3 \beta -1) (\gamma -1)}{\beta  (5 \gamma +7)-\gamma -2}}.
                \end{split}
        \end{equation}
        In order to avoid the singularities in density and pressure profiles that corresponds to the big bang singularity at $R(t)=0$, it requires  $t\neq -\frac{R_1}{R_0}$.  Combining the constraint on $\omega$ and $\beta$ in (\ref{67.1}) with $0<\beta<\frac{1}{6}$, the flaring-out, flatness and weak energy condition
        can all be satisfied simultaneously if
        \begin{equation}
                \label{72.1}
                                        \quad R_0 R_1>0,~~~~\frac{2 \beta -1}{14 \beta -4}<\gamma <\frac{16 \beta -5}{4 \beta -2},~~~~~~~~ r_0>\frac{1}{2} \sqrt{\frac{14 \beta  \gamma -2 \beta -4 \gamma +1}{R_0^2 R_1^2 (5 \beta  \gamma +7 \beta -\gamma -2)}}.
                        \end{equation} 
Here one observes that the satisfaction of all wormhole conditions imposes some interesting constraints. Specifically: $(i)$ the required matter type ($\gamma$ and $\omega$ parameters) for a specific solution is constrained by the Rastall's coupling, and $ii)$ the wormhole throat radius $r_0$ cannot be arbitrary, and it constrained by the Rastall's coupling $\beta$ and the matter parameter $\gamma$.  This is similar to the result in \cite{cjp} where it is shown that for wormholes in the Einstein-de Sitter universe, the wormhole throat radius not only depends on the shape function parameters but also on the background cosmological constant.  
 For a specific set of parameters according to the constraints (\ref{72.1}), the behavior of  $\rho$, $\rho+P_r$ and $\rho+P_l$ as well as $B(r)/r$  are illustrated in the Figures \ref{fig.3} and \ref{fig.4}. The positiveness of $\rho$, $\rho+P_r$ and $\rho+P_l$ represents the satisfaction of the WEC
in Rastall's theory. Figure \ref{fig.3} shows that  for $\beta=0.163$ with
variety of 
$\gamma$ values in the range given by \eqref{72.1}, the WEC condition remains respected
for a variety of  wormholes with radii $r_0$ satisfying \eqref{72.1}.  Here one notes that the throat radius $r_0$ is fixed for a fixed value of $\beta$ and $\gamma$, and is defined as the point where $B(r)$ is minimum. In case of a dynamic wormhole the throat area is subject to change in time  due to changing $R(t)$. In Figure \ref{fig.4}, the first plot represents the asymptotic
flatness of $B(r)/r$ function and the other plots represent the satisfaction of WEC for a specific wormhole with the characteristic
parameters $r_0=0.1, \beta=0.163, \gamma=0.4$.  \\
        \begin{figure*}[ht]
                \begin{minipage}{\textwidth}
                        \includegraphics[width=\textwidth]{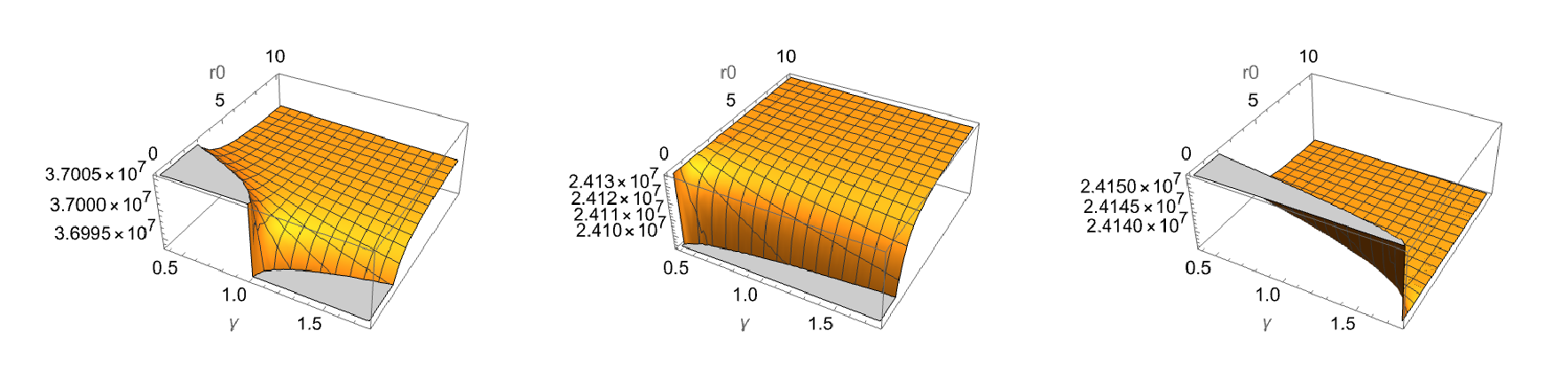}
                        \caption{\label{fig.3} This figure represents the behavior of $\rho$, $\rho+P_r$ and $\rho+P_l$, respectively, for various wormholes characterized by the throat radius $r_0$ and  $\gamma$ parameter given in \eqref{72.1}. Here we considered the set of constants $R_0 = 2,\, R_1=3,\, \beta=0.163,\, t=1 \, \mbox{and} \, G=6.67 \times 10^{-11}$ .}
                \end{minipage}
        \end{figure*}
        \begin{figure*}[ht]
                \begin{minipage}{\textwidth}
                        \includegraphics[width=\textwidth]{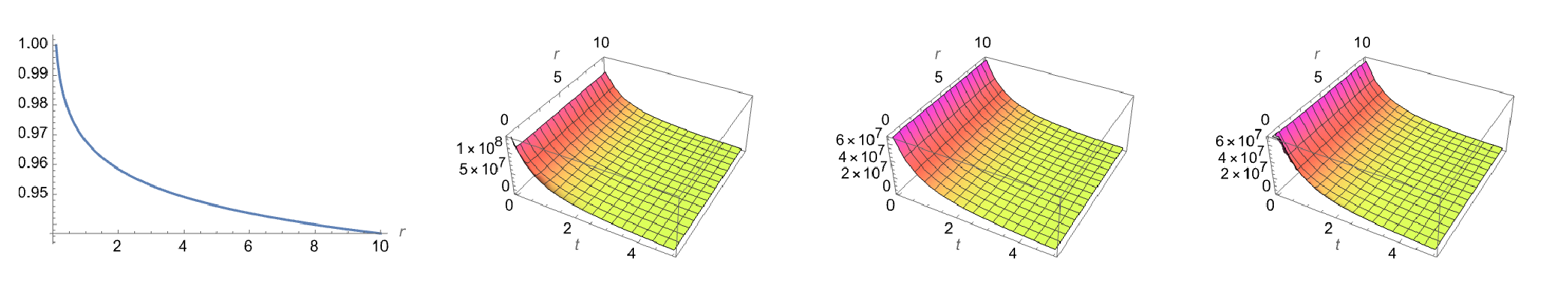}
                        \caption{\label{fig.4} This figure indicates the behavior of $\frac{B(r)}{r}$, $\rho$, $\rho+P_r$ and $\rho+P_l$ versus the radial coordinate r and time coordinate t, respectively. Here we considered the set of constants  $R_0 = 2,\, R_1=3,\, r_0=0.1,\, \beta=0.163,\, \gamma=0.4 \, \mbox{and} \, G=6.67 \times 10^{-11}$.}
                \end{minipage}
        \end{figure*}
        
        \item[$\bullet$] \label{rr3} 
        
        \textit{$\boldsymbol{a}=\boldsymbol{\frac{1}{2} d}$}\\
        
       Inserting the scale factor  and and shape function
       in (\ref{73.1}) into the field equations (\ref{17}-\ref{19}), we  find
        \begin{equation}
                \begin{split}
                        \label{75.1}
                        \rho(t,r)& =\frac{3 R_0^2 (6 \beta -1) }{32 \pi G (4 \beta -1) (R_0 t+R_1)^2}\\
                        & +\frac{3 r_0^{-2} (6 \beta -1) (2 \beta -1) (\gamma -1)}{8 \pi G  (4 \beta -1) (8 \beta  (2 \gamma +1)-5 \gamma -1) (R_0 t+R_1)} \, \left(\frac{r_0}{r}\right)^{\frac{3 (\gamma -1)}{8 \beta  (2 \gamma +1)-5 \gamma -1}+3},
                \end{split}
        \end{equation}
        
        \begin{equation}
                \begin{split}
                        \label{76.1}
                        \rho(t,r)+P_r(t,& r)= \frac{(6 \beta -1) R_0^2}{8 \pi  (4 \beta -1) G (R_0 t+R_1)^2} \\
                        & -\frac{r_0^{-2}(6 \beta -1) (\beta  (8 \gamma +4)-\gamma -2)}{4 \pi G (4 \beta -1)(8 \beta  (2 \gamma +1)-5 \gamma -1) (R_0 t+R_1)} \, \left(\frac{r_0}{r}\right)^{\frac{3 (\gamma -1)}{8 \beta  (2 \gamma +1)-5 \gamma -1}+3},
                \end{split}
        \end{equation}
        
        \begin{equation}
                \begin{split}
                        \label{77.1}
                        \rho(t,r)+P_l(t,&r)=\frac{(6 \beta -1) R_0^2}{16 \pi G (4 \beta -1) (R_0 t+R_1)^2} \\
                        &+\frac{r_0^{-2}(6 \beta -1) (\beta  (8 \gamma +4)-4 \gamma +1)}{16 \pi  G (4 \beta -1) (8 \beta  (2 \gamma +1)-5 \gamma -1) (R_0 t+R_1)} \, \left(\frac{r_0}{r}\right)^{\frac{3 (\gamma -1)}{8 \beta  (2 \gamma +1)-5 \gamma -1}+3}.
                \end{split}
        \end{equation}
        In this case, satisfaction of flaring-out condition, flatness condition and WEC at throat requires
        \begin{equation}
                \label{78.1}
                \begin{split}
                        &R_0,R_1<0: 
                \end{split}
                \begin{cases}
                        \gamma <\frac{2-4 \beta }{8 \beta -1}, ~~ 0<\beta<\frac{1}{8}; &~~~~ r_0\geq 2 \sqrt{-\frac{2 \beta  \gamma  R_1-2 \beta  R_1-\gamma  R_1+R_1}{R_0^2 (16 \beta \gamma +8 \beta -5 \gamma -1)}}, \\
                        \gamma >\frac{-4 \beta -1}{8 \beta -4},~~ 0<\beta<\frac{1}{8}; &~~~~ r_0>\sqrt{\frac{-8 \beta  \gamma  R_1-4 \beta  R_1+4 \gamma  R_1-R_1}{R_0^2 (16 \beta  \gamma +8 \beta -5 \gamma -1)}},  \\
                        \gamma >\frac{1}{2},~~ \beta =\frac{1}{8} ; &~~~~ r_0>\sqrt{-\frac{6 \gamma  R_1-3 R_1}{6 \gamma  R_0^2}},\\
                        \frac{-4 \beta -1}{8 \beta -4}<\gamma <\frac{2-4 \beta }{8 \beta -1}, ~~~~ \frac{1}{8}<\beta < \frac{1}{6}; & ~~~~ r_0>\sqrt{\frac{-8 \beta  \gamma  R_1-4 \beta  R_1+4 \gamma  R_1-R_1}{R_0^2 (16 \beta  \gamma +8 \beta -5 \gamma -1)}} .
                \end{cases}
        \end{equation}
        
        \begin{equation}
                \label{78.2}   
                \begin{split}
                        &R_0,R_1>0:
                \end{split}
                \begin{cases}
                        \gamma <\frac{2-4 \beta }{8 \beta -1},~~\mbox{or}~~ \gamma >\frac{-4 \beta -1}{8 \beta -4}; ~~0<\beta<\frac{1}{8}; &r_0>\sqrt{2} \sqrt{\frac{8 \beta  \gamma  R_1+4 \beta  R_1-\gamma R_1-2 R_1}{R_0^2 (16 \beta  \gamma +8 \beta -5 \gamma -1)}},\\
                        \gamma >\frac{1}{2}, ~~\beta =\frac{1}{8};& r_0>\sqrt{\frac{R_1}{\gamma R_0^2}}, \\
                        \frac{-4 \beta -1}{8 \beta -4}<\gamma <\frac{2-4 \beta }{8 \beta -1}, ~~\frac{1}{8}<\beta < \frac{1}{6}; &r_0>\sqrt{2} \sqrt{\frac{8 \beta  \gamma R_1+4 \beta  R_1-\gamma  R_1-2 R_1}{R_0^2 (16 \beta  \gamma +8 \beta -5 \gamma -1)}}.
                \end{cases}
        \end{equation}
       Similar arguments given for the previous solution and its figures can be also made here. Figure \ref{fig.5} shows that for a specific $\beta=0.041$, the WEC will be satisfied for variety of wormholes with $r_0$ and $\gamma$ meeting the constraints in \eqref{78.1}.
                Figure \ref{fig.6} shows the asymptotic behavior of $B(r)$,
                as well as   $\rho$, $\rho+P_r$ and $\rho+P_l$ satisfying the WECs for a specific set of parameters according to the constraints (\ref{78.1}) in the entire spacetime.
        \begin{figure*}[ht]
                \begin{minipage}{\textwidth}
                        \includegraphics[width=\textwidth]{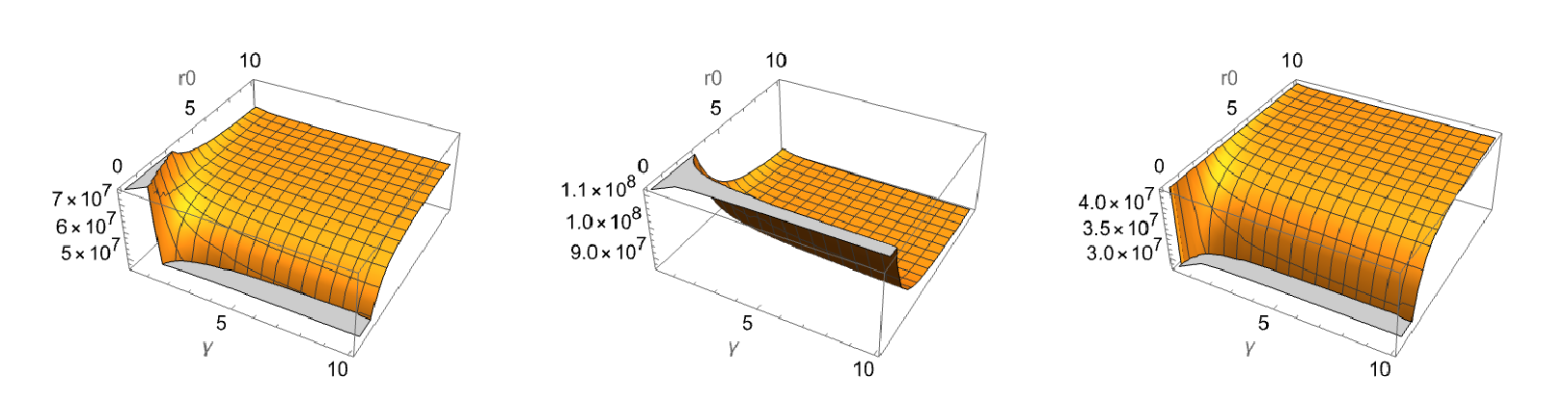}
                        \caption{\label{fig.5}This figure represents the behavior of $\rho$, $\rho+P_r$ and $\rho+P_l$, respectively, for various wormholes characterized by the throat radius $r_0$ and  $\gamma$ parameter given in \eqref{78.1}. Here we considered the set of constants $R_0 =-2,\, R_1=-3,\, \beta=0.041,\, t=1  \, \mbox{and} \, G=6.67 \times 10^{-11}$.}
                \end{minipage}
        \end{figure*}
        \begin{figure*}[ht]
                \begin{minipage}{\textwidth}
                        \includegraphics[width=\textwidth]{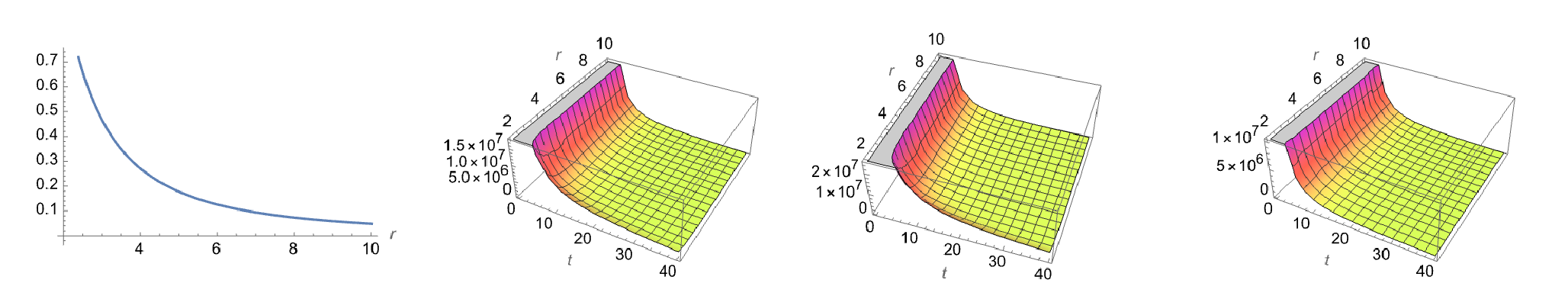}
                        \caption{\label{fig.6} This figure represents the behavior of $\frac{B(r)}{r}$, $\rho$, $\rho+P_r$ and $\rho+P_l$, respectively, for various wormholes characterized by throat radius $r_0$ and  $\gamma$ parameter given in \eqref{78.1}. Here we considered the set of constants $R_0 =-2,\, R_1=-3,\,r_0=2,\, \beta=0.041,\, \omega=3,\, \gamma=0.35 \, \mbox{and} \, G=6.67 \times 10^{-11}$ .}
                \end{minipage}
        \end{figure*}
\end{itemize}
\subsubsection{\label{wec>0-cne0} Analysis of  solutions for $C\ne 0$}
\begin{figure*}[ht]
        \begin{minipage}{\textwidth}
                \includegraphics[width=\textwidth]{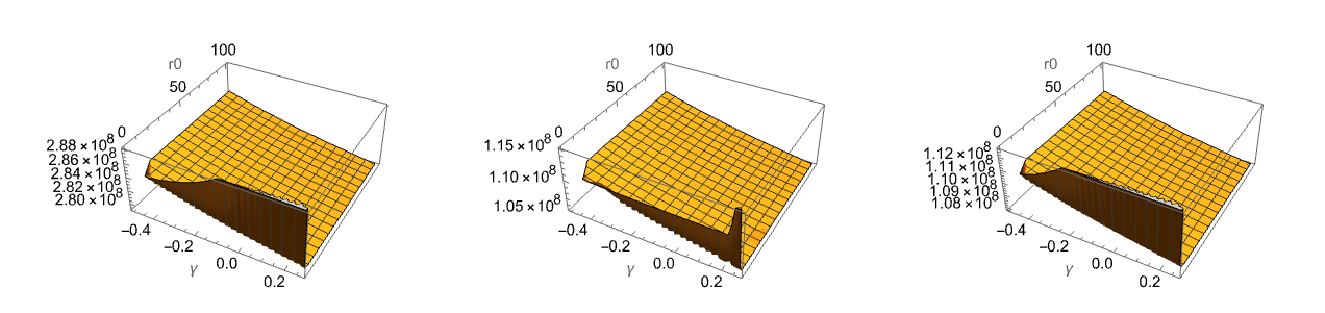}
                \caption{\label{fig.7} This figure represents the behavior of $\rho$, $\rho+P_r$ and $\rho+P_l$, respectively, for various wormholes characterized by the  throat radius $r_0$ and  $\gamma$ parameter given in \eqref{113}. Here we considered the set of constants  $R_1=5,\, R_0 = 2,\, C=-3,\, \beta=0.041,\, t=1 \, \, \mbox{and} \, G=6.67 \times 10^{-11}$.}
        \end{minipage}
\end{figure*}

\begin{figure*}[ht]
        \begin{minipage}{\textwidth}
                \includegraphics[width=\textwidth]{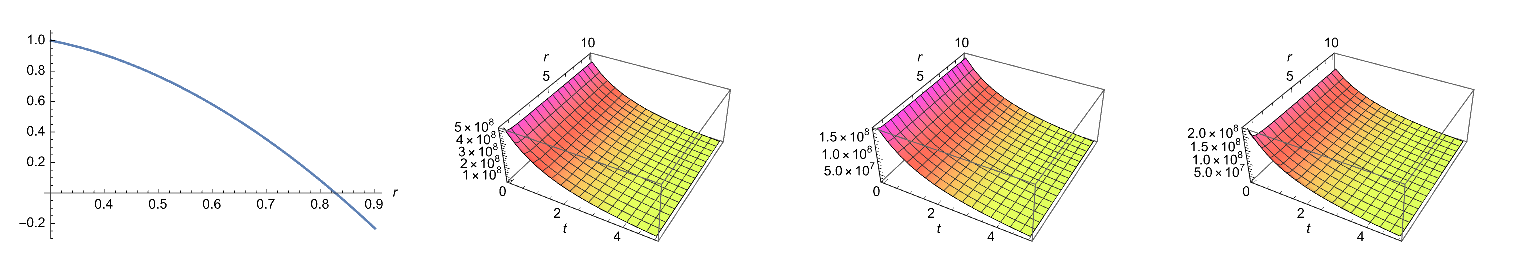}
                \caption{\label{fig.8}This figure indicates the behavior of $\frac{B(r)}{r}$, $\rho$, $\rho+P_r$ and $\rho+P_l$ regarding r and t from left to right respectively for $0.3<r<10 $ and $0<t<5 $. The arbitrary constants are taken as $R_0 = 2,\,  \beta=0.041,\, C=-3,\, R_1=5,\, r_0=0.3,\, \gamma=0.2 \, \,\mbox{and} \, G=6.67 \times 10^{-11}$ .}
        \end{minipage}
\end{figure*}

Here we analyze the following two particular cases.
\begin{itemize}
        \item[$\bullet$] \textit{ $\mathbf{a=-2 b}$}\\
        
       Substituting the scale factor and shape function in (\ref{101})  into the field equations (\ref{17}-\ref{19}), we find
\end{itemize}
        \begin{equation}
        	\small
                \label{110}
                \begin{split}
                        \rho(t,r)&=\frac{3 C R_0^2 (3 \beta -1) (6 \beta -1) (9 \beta -2) }{2 \pi G (4 \beta -1) \left(-48 \beta +R_0^2 \left(2 \sqrt{3} (4 \beta -1) R_1 t \sqrt{\frac{(2-9 \beta ) C}{4 \beta -1}}+(2-9 \beta ) C t^2+3 (4 \beta -1) R_1^2\right)+12\right)}\\
                        &-\frac{162 R_0^2 (2 \beta -1) (3 \beta -1) (6 \beta -1) (\gamma -1)  \left(C (9 \beta -2) +r_0^{-2}(3-12 \beta))\right)}{\pi G (1-4 \beta )^2 (\beta  (5 \gamma +7)-\gamma -2) \left(R_0^2 \left(\sqrt{3} t \sqrt{\frac{(2-9 \beta ) C}{4 \beta -1}}+3 R_1\right)^2-36\right)^2} \, \left(\frac{r_0}{r}\right)^{3-\frac{3 (3 \beta -1) (\gamma -1)}{-\beta  (5 \gamma +7)+\gamma +2}},
                \end{split}
        \end{equation}
        \begin{equation}
        	\small
                \label{111}
                \begin{split}
                        \rho&(t,r)+P_r(t,r)=\frac{C R_0^2 (6 \beta -1) (9 \beta -2) }{2 \pi G (4 \beta -1) \left(48 \beta +R_0^2 \left(2 \sqrt{3} (1-4 \beta ) R_1 t \sqrt{\frac{(2-9 \beta ) C}{4 \beta -1}}+(9 \beta -2) C t^2+(3-12 \beta ) R_1^2\right)-12\right)} \\
                        &+ \frac{6 R_0^2 (6 \beta -1)  (2 \beta  (7 \gamma -1)-4 \gamma +1) \left(C(9 \beta -2)+r_0^{-2}(3-12 \beta)\right)}{\pi  G (\beta  (5 \gamma +7)-\gamma -2) \left(-48 \beta +R_0^2 \left(2 \sqrt{3} (4 \beta -1) R_1 t \sqrt{\frac{(2-9 \beta ) C}{4 \beta -1}}+(2-9 \beta ) C t^2+3 (4 \beta -1) R_1^2\right)+12\right)^2}\\
                        &\, \left(\frac{r_0}{r}\right)^{3-\frac{3 (3 \beta -1) (\gamma -1)}{-\beta  (5 \gamma +7)+\gamma +2}},
                \end{split}
        \end{equation}
        \begin{equation}
        	\small
                \label{112}
                \begin{split}
                        \rho&(t,r)+P_l(t,r)=\frac{C R_0^2 (6 \beta -1) (9 \beta -2)}{2 \pi G (4 \beta -1) \left(48 \beta +R_0^2 \left(2 \sqrt{3} (1-4 \beta ) R_1 t \sqrt{\frac{(2-9 \beta ) C}{4 \beta -1}}+(9 \beta -2) C t^2+(3-12 \beta ) R_1^2\right)-12\right)}\\
                        &+ \frac{3 R_0^2 (6 \beta -1)  (4 \beta  (\gamma -4)-2 \gamma +5) \left(C(9 \beta -2)+r_0^{-2}(3-12 \beta))\right)}{\pi  G (\beta  (5 \gamma +7)-\gamma -2) \left(-48 \beta +R_0^2 \left(2 \sqrt{3} (4 \beta -1) R_1 t \sqrt{\frac{(2-9 \beta ) C}{4 \beta -1}}+(2-9 \beta ) C t^2+3 (4 \beta -1) R_1^2\right)+12\right)^2}\\
                        &\, \left(\frac{r_0}{r}\right)^{3-\frac{3 (3 \beta -1) (\gamma -1)}{-\beta  (5 \gamma +7)+\gamma +2}}.
                \end{split}
        \end{equation}
        Since $0<\beta<\frac{1}{6}$ and $\omega =\frac{3-9 \beta }{9 \beta -2}$,  the WEC and flaring-out condition will be satisfied under the fallowing conditions 
        \begin{equation}
                \label{113}
                C<0:
                \begin{cases}
                        R_0\leq \frac{-2}{\left|R_1\right|}, ~~ \mbox{or} ~~ R_0>\frac{2}{\left|R_1\right|}; ~~ -\frac{1}{2}<\gamma \leq \frac{2 \beta -1}{14 \beta -4}, ~~~~~~~~~~  & r_0>\sqrt{\frac{14 \beta  \gamma -2 \beta -4 \gamma +1}{(9 \beta -2) (2 \gamma +1) C}},\\
                        \frac{-2}{\left|R_1\right|}<R_0<\frac{2}{\left|R_1\right|}, ~~ -\frac{1}{2}<\gamma \leq \frac{20 \beta -7 \beta  R_0^2 R_1^2+2 R_0^2 R_1^2-4}{-76 \beta +5 \beta  R_0^2 R_1^2-R_0^2 R_1^2+20}, ~~~~~
~  & r_0>\sqrt{\frac{14 \beta  \gamma -2 \beta -4 \gamma +1}{(9 \beta -2) (2 \gamma +1) C}}.  
                \end{cases}
        \end{equation}
\\
Similar arguments given for the previous solutions and their figures can be also made here. Figure \ref{fig.7} shows that for a specific $\beta=0.041$, the WEC will be satisfied for variety of wormholes with $r_0$ and $\gamma$ meeting the constraints in  (\ref{113}). Figure \ref{fig.8} shows the behavior of $B(r)/r$ as well as $\rho$, $\rho+P_r$ and $\rho+P_l$ satisfying the WEC for a specific set of parameters according to the constraints in  (\ref{113}). As it is seen from the first plot, in this case
we have a finite wormhole configuration which cannot be arbitrarily large.


\begin{figure*}[ht]
        \begin{minipage}{\textwidth}
                \includegraphics[width=\textwidth]{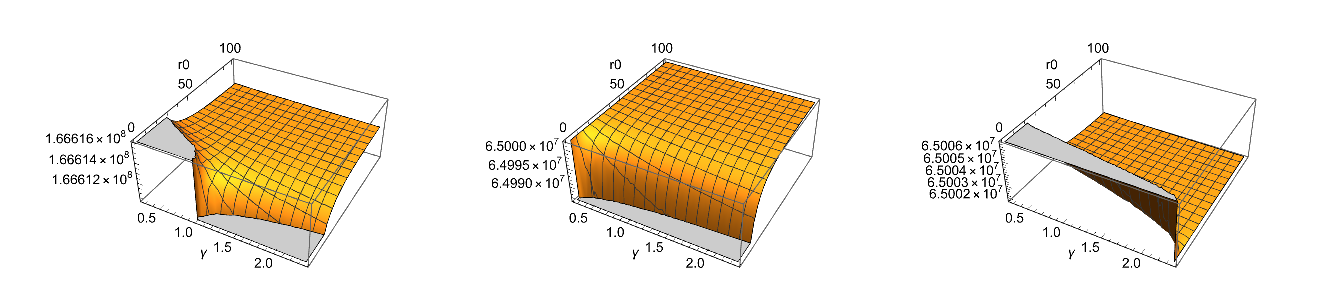}
                \caption{\label{fig.9} This figure represents the behavior of $\rho$, $\rho+P_r$ and $\rho+P_l$, respectively, for various wormholes characterized by the throat radius $r_0$ and  $\gamma$ parameter given in \eqref{113-2}. Here we considered the set of constants  $R_1=5,\, R_0 = 2,\, \beta=0.041,\, t=1 \, \, \mbox{and} \, G=6.67 \times 10^{-11}$.}
        \end{minipage}
\end{figure*}

\begin{figure*}[ht]
        \begin{minipage}{\textwidth}
                \includegraphics[width=\textwidth]{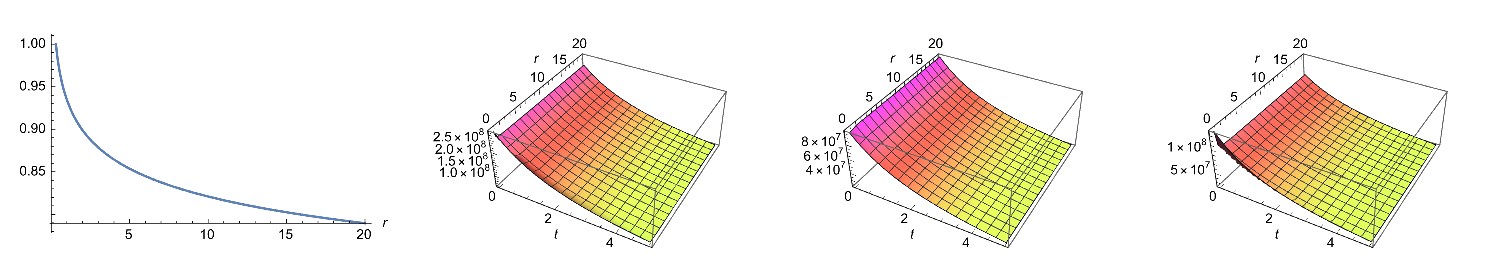}
                \caption{\label{fig.10} This figure indicates the behavior of $\frac{B_n(r)}{r}$, $\rho$, $\rho+P_r$ and $\rho+P_l$ regarding r and t from left to right respectively for $0.3<r<20 $ and $0<t<5 $. The arbitrary constants are taken as $R_0 = 2,\,  \beta=0.041,\, R_1=5,\, r_0=0.3,\, \gamma=0.3 \, \,\mbox{and} \, G=6.67 \times 10^{-11}$ .}
        \end{minipage}
\end{figure*}

\begin{itemize}
        \item[$\bullet$] \textit{ $\mathbf{a=-2 b, \,\, k=1}$}\\
        
         Considering the shape function and scale factor as \eqref{101-2} leaves the field equations (\ref{17}-\ref{19}) as
         
         \begin{equation}
                \label{110-2}
                \begin{split}
                        \rho(t,r)&=-\frac{3 (6 \beta -1) R_0^2 \left((3 \beta -1) R_0^2 (R_1 \pm t)^2-4 \beta \right)}{2 \pi  (4 \beta -1) G \left(R_0^2 (R_1 \pm t)^2-4\right)^2}\\
                        &+\frac{6 (6 \beta -1) (\beta  (6 \beta -5)+1) (\gamma -1) R_0^2 r_0^{-2}}{ \left(G \pi  (4 \beta -1) (\beta  (5 \gamma +7)-\gamma -2) \left(R_0^2 (R_1 \pm t)^2-4\right)^2\right)} \, \left(\frac{r_0}{r}\right)^{3-\frac{3 (3 \beta -1) (\gamma -1)}{-\beta  (5 \gamma +7)+\gamma +2}},
                \end{split}
         \end{equation}\\
         \begin{equation}
                \label{111-2}
                \begin{split}
                        \rho(t,r)+P_r(t,r)&=\frac{(6 \beta -1) R_0^2 \left(R_0^2 (R_1 \pm t)^2+4\right)}{2 \pi G (4 \beta -1) \left(R_0^2 (R_1 \pm t)^2-4\right)^2} \\
                        &- \frac{2 R_0^2 r_0^{-2}(6 \beta -1) (2 \beta  (7 \gamma -1)-4 \gamma +1)}{\pi G (4 \beta -1) (\beta  (5 \gamma +7)-\gamma -2) \left(R_0^2 (R_1\pm t)^2-4\right)^2} \, \left(\frac{r_0}{r}\right)^{3-\frac{3 (3 \beta -1) (\gamma -1)}{-\beta  (5 \gamma +7)+\gamma +2}},
                \end{split}
         \end{equation}\\
         \begin{equation}
                \label{112-2}
                \begin{split}
                        \rho(t,r)+P_l(t,r)&=\frac{(6 \beta -1) R_0^2 \left(R_0^2 (R_1\pm t)^2+4\right)}{2  \pi G (4 \beta -1)\left(R_0^2 (R_1\pm t)^2-4\right)^2}\\
                        &- \frac{r_0^{-2} R_0^2 (6 \beta -1) (4 \beta  (\gamma -4)-2 \gamma +5)}{ \left(\pi G (4 \beta -1) G (\beta  (5 \gamma +7)-\gamma -2) \left(R_0^2 (R_1 \pm t)^2-4\right)^2\right)} \, \left(\frac{r_0}{r}\right)^{3-\frac{3 (3 \beta -1) (\gamma -1)}{-\beta  (5 \gamma +7)+\gamma +2}}.
                \end{split}
         \end{equation}
\end{itemize}
Considering $0<\beta<\frac{1}{6}$ and $\omega=\frac{3-9 \beta}{9 \beta -2}$, the WEC, flaring-out and asymptotically flantess condition will be satisfied simultaneously if
\begin{equation}
        \label{113-2}
        R_0\ne0,~ R_0\ne \frac{\pm 2}{|R_1|}, ~~~~  \frac{2 \beta -1}{14 \beta -4}<\gamma <\frac{16 \beta -5}{4 \beta -2}, ~~~~ r_0>2 \sqrt{\frac{14 \beta  \gamma -2 \beta -4 \gamma +1}{(5 \beta  \gamma +7 \beta -\gamma -2) \left(R_0^2 R_1^2+4\right)}}
\end{equation}
Figure \ref{fig.9} shows that the WEC will be satisfied for variety of wormholes with $r_0$ and $\gamma$ meeting the constraints in \eqref{113-2}. Figure \ref{fig.10} shows the asymptotic behavior of $B_n(r)/r$ as well as $\rho$, $\rho+P_r$, and $\rho+P_l$ respecting WEC in entire spacetime. 
\subsection{\label{beta<0} WEC for $\beta<0$}
Some observational tests of Rastall theory indicates negative values of $\beta$~,
see as an instance  \cite{PRD-moradpour}. Hence, in this subsection we address the WEC and flaring-out condition for $\beta<0$.
\subsubsection{\label{wec<0-c=0} Analysis of solutions for $C=0$}
\begin{itemize}
        \item[$\bullet$] \label{rrr2} \textit{ $\mathbf{a=2d}$}\\
        
        In order to satisfy the WEC, flaring-out condition and flatness condition in this case, using Eq.(\ref{67.1}) with $\beta <0$, the following constraints should be satisfied. 
        \begin{equation}
                \label{72.2}
                                          \quad R_0 R_1>0,~~~~\frac{2 \beta -1}{14 \beta -4}<\gamma <\frac{16 \beta -5}{4 \beta -2},~~~~~~~~~~ r_0>\frac{1}{2} \sqrt{\frac{14 \beta  \gamma -2 \beta -4 \gamma +1}{R_0^2 R_1^2 (5 \beta  \gamma +7 \beta -\gamma -2)}}.
                        \end{equation} 
        The constraints here are the same as the obtained ones for $0<\beta<\frac{1}{6}$ in (\ref{72.1}).

        \item[$\bullet$] \label{rrr3} \textit{ $\mathbf{a=\frac{1}{2}d}$}\\
        
        Using  (\ref{73.1}), since $\omega=3$ and $\beta<0$, the following restrictions on $\gamma$ parameter provides  respecting the WE, flaring-out and flatness conditions
        \begin{equation}
                \label{116}
                \begin{split}
                        &R_0, R_1<0: 
                \end{split}
                \begin{cases}
                        \gamma <\frac{2-4 \beta }{8 \beta -1},~~~~~ &r_0\geq 2 \sqrt{-\frac{2 \beta  \gamma  R_1-2 \beta  R_1-\gamma  R_1+R_1}{R_0^2 (16 \beta  \gamma +8 \beta -5 \gamma -1)}}, \\
                        \gamma >\frac{-4 \beta -1}{8 \beta -4},~~~~~~  & r_0>\sqrt{\frac{-8 \beta  \gamma  R_1-4 \beta R_1+4 \gamma  R_1-R_1}{R_0^2 (16 \beta  \gamma +8 \beta -5 \gamma -1)}}.
                \end{cases}
        \end{equation}
        \begin{equation}
                \label{116.2}   
                \begin{split}
                        &R_0, R_1>0: 
                \end{split}
                \begin{cases}
                        \gamma <\frac{2-4 \beta }{8 \beta -1}, ~~ \mbox{or}~~ \gamma \geq 0; ~~ \beta <-\frac{1}{4}; &r_0>\sqrt{2} \sqrt{\frac{8 \beta  \gamma R_1+4 \beta  R_1-\gamma R_1-2 R_1}{R_0^2 (16 \beta  \gamma +8 \beta -5 \gamma -1)}},\\
                        \frac{-4 \beta -1}{8 \beta -4}<\gamma <0, ~~\beta <-\frac{1}{4}; & r_0 \geq 2 \sqrt{-\frac{2 \beta  \gamma  R_1-2 \beta R_1-\gamma  R_1+R_1}{R_0^2 (16 \beta  \gamma +8 \beta -5 \gamma -1)}},  \\
                        \gamma <\frac{2-4 \beta }{8 \beta -1},~~ \mbox{or} ~~ \gamma >\frac{-4 \beta -1}{8 \beta -4}; ~~  -\frac{1}{4}\leq \beta <0; &r_0>\sqrt{2} \sqrt{\frac{8 \beta  \gamma R_1+4 \beta  R_1-\gamma R_1-2 R_1}{R_0^2 (16 \beta  \gamma +8 \beta -5 \gamma -1)}}.    
                \end{cases}
        \end{equation}
\end{itemize}
\subsubsection{\label{wec<0-cne0} For solution of $C\ne 0$}
\begin{itemize}
        \item[$\bullet$] \textit{ $\mathbf{a=-2b}$}\\
        
        Considering Eq.\ref{101}, the WEC and flaring-out condition will be met if
\end{itemize}
        \begin{equation}
                \label{117}
                C<0:
                \begin{cases}
                        R_0\leq \frac{-2 \sqrt{5}}{\left|R_1\right|} ~~ \mbox{or} ~~ R_0>\frac{2 \sqrt{5}}{\left|R_1\right|}, ~~ -\frac{1}{2}<\gamma \leq \frac{2 \beta -1}{14 \beta -4},  &~~~~ r_0>\sqrt{\frac{14 \beta  \gamma -2 \beta -4 \gamma +1}{(9 \beta -2) (2 \gamma +1) C}},\\
                        \frac{-2 \sqrt{5}}{\left|R_1\right|}<R_0<\frac{2 \sqrt{5}}{\left|R_1\right|}, ~~ -\frac{1}{2}<\gamma \leq \frac{20 \beta -7 \beta  R_0^2 R_1^2+2 R_0^2 R_1^2-4}{-76 \beta +5 \beta  R_0^2 R_1^2-R_0^2 R_1^2+20}, &~~~~ r_0>\sqrt{\frac{14 \beta  \gamma -2 \beta -4 \gamma +1}{(9 \beta -2) (2 \gamma +1) C}}.  
                \end{cases}
        \end{equation}

 \begin{itemize}
        \item[$\bullet$] \textit{ $\mathbf{a=-2b, k=1}$}\\
       
        Considering \eqref{101-2}, with $\beta<0$ and $\omega=\frac{3-9 \beta}{9 \beta-2}$, the WEC, flaring-out, and flatness conditions will be respected if
  \begin{equation}
        \label{117-2}
        R_0\ne0,~ R_0\ne \frac{\pm 2}{|R_1|}, ~~~~  \frac{2 \beta -1}{14 \beta -4}<\gamma <\frac{16 \beta -5}{4 \beta -2}, ~~~~ r_0>2 \sqrt{\frac{14 \beta  \gamma -2 \beta -4 \gamma +1}{(5 \beta  \gamma +7 \beta -\gamma -2) \left(R_0^2 R_1^2+4\right)}}
 \end{equation}
\end{itemize} 

\section{Conclusion} 
In this paper, analytical evolving wormhole solutions with a constant redshift function are investigated in the context of Rastall's modified theory. A general class of solutions, including the asymptotically flat and (anti)de Sitter solutions, is derived by assuming a particular equation of state for the energy density and pressure profiles. Regarding the theoretical and observational constraints on Rastall's coupling $\beta$, two admissible ranges $0<\beta<\frac{1}{6}$ and $\beta<0$ are considered in order to study the solutions versus the required
conditions for traversable wormholes. It is shown that simultaneous satisfaction of all these conditions is achievable under the obtained constraints on the parameters of the solutions. Also it is shown that the size of the wormhole throat is constrained and depends on both the Rastall's coupling $\beta$ and the equation of state parameters of the matter source. A list of three particular solutions with their constraints providing the satisfaction of all wormhole conditions is given in Table \ref{tab1}.
\renewcommand{\arraystretch}{3}
\FloatBarrier
\begin{table*}[ht]
        \centering
        \resizebox{\textwidth}{!}{\begin{tabular}{ |c|c|c|c|c|c|c| }
                        \hline
                        $C$ & Metric functions & $\omega$ & $\beta$ & $\gamma$& $r_0$& $R_0$, $R_1$\\
                        \hline
                        \multirow{14}{3em}{$C=0$} & \multirow{2}{10em}{\centering \makecell{$R(t)=( R_0 \,t +  R_1)^2$ \\ \\ $B(r)=r_0 \left(\frac{r_0}{r}\right)^{\frac{3 (3 \beta -1) (\gamma -1)}{\beta  (5 \gamma +7)-\gamma -2}}$}} & \multirow{2}{5em}{\centering$\omega =\frac{9 \beta -3}{2-9 \beta }$} &$0< \beta< \frac{1}{6}$ & $\frac{2 \beta -1}{14 \beta -4}<\gamma <\frac{16 \beta -5}{4 \beta -2}$ & $r_0>\frac{1}{2} \sqrt{\frac{14 \beta  \gamma -2 \beta -4 \gamma +1}{R_0^2 R_1^2 (5 \beta  \gamma +7 \beta -\gamma -2)}} $& \multirow{2}{5em}{\centering $R_0 R_1 >0$}\\
                        \cline{4-6}
                        & & & $\beta <0 $& $\frac{2 \beta -1}{14 \beta -4}<\gamma <\frac{16 \beta -5}{4 \beta -2}$ &$ r_0>\frac{1}{2} \sqrt{\frac{14 \beta  \gamma -2 \beta -4 \gamma +1}{R_0^2 R_1^2 (5 \beta  \gamma +7 \beta -\gamma -2)}} $& \\
                        \cline{2-7}
                        & \multirow{12}{10em}{\centering \makecell{$R(t)=( R_0 \,t+ R_1)^{\frac{1}{2}}$\\ \\ $B(r)=r_0 \left(\frac{r_0}{r}\right)^{\frac{3 (\gamma -1)}{8 \beta  (2 \gamma +1)-5 \gamma -1}}$}} &\multirow{12}{5em}{\centering $\omega =3$} &\multirow{3}{5em}{\centering $0< \beta < \frac{1}{8}$} & $\gamma <\frac{2-4 \beta }{8 \beta -1} $ &  $r_0\geq 2 \sqrt{-\frac{2 \beta  \gamma  R_1-2 \beta  R_1-\gamma  R_1+R_1}{R_0^2 (16 \beta \gamma +8 \beta -5 \gamma -1)}} $ &\multirow{2}{5em}{\centering \makecell{ $R_0<0,$\\ $R_1<0$}} \\
                        \cline{5-6}
                        & & & &$ \gamma >\frac{-4 \beta -1}{8 \beta -4}$& $r_0>\sqrt{\frac{-8 \beta  \gamma  R_1-4 \beta  R_1+4 \gamma  R_1-R_1}{R_0^2 (16 \beta  \gamma +8 \beta -5 \gamma -1)}}$& \\
                        \cline{5-7}
                        & & & & $\gamma <\frac{2-4 \beta }{8 \beta -1} \, \, \mbox{or} \, \,  \gamma >\frac{-4 \beta -1}{8 \beta -4}$& $r_0>\sqrt{2} \sqrt{\frac{8 \beta  \gamma  R_1+4 \beta  R_1-\gamma R_1-2 R_1}{R_0^2 (16 \beta  \gamma +8 \beta -5 \gamma -1)}}$& \makecell{ $R_0>0,$\\ $R_1>0$}\\
                        \cline{4-7}
                        & & & \multirow{2}{5em}{\centering $\beta=\frac{1}{8}$}& \multirow{2}{5em}{$\gamma > \frac{1}{2}$}&$r_0>\sqrt{-\frac{6 \gamma  R_1-3 R_1}{6 \gamma  R_0^2}}$ & \makecell{ $R_0<0,$\\ $R_1<0$} \\
                        \cline{6-7}
                        & & & & & $r_0>\sqrt{\frac{R_1}{\gamma R_0^2}}$&\makecell{ $R_0>0,$\\ $R_1>0$}\\
                        \cline{4-7}
                        & & & \multirow{2}{5em}{$\frac{1}{8}< \beta< \frac{1}{6}$}& \multirow{2}{9em}{$\frac{-4 \beta -1}{8 \beta -4}<\gamma <\frac{2-4 \beta }{8 \beta -1}$}& $r_0>\sqrt{\frac{-8 \beta  \gamma  R_1-4 \beta  R_1+4 \gamma  R_1-R_1}{R_0^2 (16 \beta  \gamma +8 \beta -5 \gamma -1)}}$ &\makecell{ $R_0<0,$\\ $R_1<0$} \\
                        \cline{6-7}
                        & & & & &$r_0>\sqrt{2} \sqrt{\frac{8 \beta  \gamma R_1+4 \beta  R_1-\gamma  R_1-2 R_1}{R_0^2 (16 \beta  \gamma +8 \beta -5 \gamma -1)}}$&\makecell{ $R_0>0,$\\ $R_1>0$}\\
                        \cline{4-7}
                        &  & & \multirow{2}{5em}{$\beta < 0$ }& $\gamma <\frac{2-4 \beta }{8 \beta -1}$ & $r_0\geq 2 \sqrt{-\frac{2 \beta  \gamma  R_1-2 \beta  R_1-\gamma  R_1+R_1}{R_0^2 (16 \beta  \gamma +8 \beta -5 \gamma -1)}}$ & \multirow{2}{5em}{\makecell{ $R_0<0,$\\ $R_1<0$}} \\
                        \cline{5-6}
                        & & & &$\gamma >\frac{-4 \beta -1}{8 \beta -4}$& $r_0>\sqrt{\frac{-8 \beta  \gamma  R_1-4 \beta R_1+4 \gamma  R_1-R_1}{R_0^2 (16 \beta  \gamma +8 \beta -5 \gamma -1)}}$& \\
                        \cline{4-7}
                        & & & \multirow{2}{5em}{$\beta<-\frac{1}{4}$} & $\gamma <\frac{2-4 \beta }{8 \beta -1} ~\mbox{or}~ \gamma \geq 0$ & $r_0>\sqrt{2} \sqrt{\frac{8 \beta  \gamma R_1+4 \beta  R_1-\gamma R_1-2 R_1}{R_0^2 (16 \beta  \gamma +8 \beta -5 \gamma -1)}}$& \multirow{3}{5em}{\makecell{ $R_0>0,$\\ $R_1>0$}}\\
                        \cline{5-6}
                        & & & & $\frac{-4 \beta -1}{8 \beta -4}<\gamma <0 $ & $r_0 \geq 2 \sqrt{-\frac{2 \beta  \gamma  R_1-2 \beta R_1-\gamma  R_1+R_1}{R_0^2 (16 \beta  \gamma +8 \beta -5 \gamma -1)}}$ & \\
                        \cline{4-6}
                        & & & $-\frac{1}{4} \leq \beta<0$& $\gamma <\frac{2-4 \beta }{8 \beta -1} ~\mbox{or}~ \gamma >\frac{-4 \beta -1}{8 \beta -4} $ & $r_0>\sqrt{2} \sqrt{\frac{8 \beta  \gamma R_1+4 \beta  R_1-\gamma R_1-2 R_1}{R_0^2 (16 \beta  \gamma +8 \beta -5 \gamma -1)}}$ & \\
                        \hline
                        \multirow{6}{3em}{$C\ne 0$ \\ and $C<0$} & \multirow{6}{13em}{ \centering \makecell{ $R(t)=\frac{1}{R_0}-\frac{R_0}{4}\left(\pm \sqrt{\frac{C}{b}}
                                        t + R_1\right)^2$ \\ \\ $B(r) =\frac{(9 \beta -2) C r^3}{12 \beta -3} $\\ $-\frac{r_0 \left((3-12 \beta )+(9 \beta -2) C r_0^2\right)}{12 \beta -3} $\\ $\left(\frac{r}{r_0}\right)^{\frac{3 (3 \beta -1) (\gamma -1)}{-\beta  (5 \gamma +7)+\gamma +2}}$}} & \multirow{6}{5em}{\centering $\omega =\frac{3-9 \beta}{9 \beta -2 }$ }& \multirow{3}{5em}{$0<\beta<\frac{1}{6}$} & \multirow{2}{10em}{\centering \makecell{ $-\frac{1}{2}<\gamma \leq $\\ \\$\frac{20 \beta -7 \beta  R_0^2 R_1^2+2 R_0^2 R_1^2-4}{-76 \beta +5 \beta  R_0^2 R_1^2-R_0^2 R_1^2+20}$}}& \multirow{2}{10em}{$r_0>\sqrt{\frac{14 \beta  \gamma -2 \beta -4 \gamma +1}{(9 \beta -2) (2 \gamma +1) C}}$} & \multirow{2}{5em}{$\frac{-2}{\left|R_1\right|}<R_0<\frac{2}{\left|R_1\right|}$} \\ 
                        & & & & & & \\
                        \cline{5-7}
                        & & & & $-\frac{1}{2}<\gamma \leq \frac{2 \beta -1}{14 \beta -4}$& $r_0>\sqrt{\frac{14 \beta  \gamma -2 \beta -4 \gamma +1}{(9 \beta -2) (2 \gamma +1) C}}$& \makecell{$R_0\leq \frac{-2}{\left|R_1\right|}$\\$\mbox{or} $\\ $R_0>\frac{2}{\left|R_1\right|}$} \\
                        \cline{4-7}
                        &  &  & \multirow{3}{4em}{\centering $\beta < 0$}  & \multirow{2}{10em}{\makecell{ $-\frac{1}{2}<\gamma \leq $\\ \\$\frac{20 \beta -7 \beta  R_0^2 R_1^2+2 R_0^2 R_1^2-4}{-76 \beta +5 \beta  R_0^2 R_1^2-R_0^2 R_1^2+20}$} } & \multirow{2}{10em}{$r_0>\sqrt{\frac{14 \beta  \gamma -2 \beta -4 \gamma +1}{(9 \beta -2) (2 \gamma +1) C}}$}& \multirow{2}{5em}{$\frac{-2 \sqrt{5}}{\left|R_1\right|}<R_0 <\frac{2 \sqrt{5}}{\left|R_1\right|}$} \\ 
                        & & & & & & \\
                        \cline{5-7}
                        & & & & $-\frac{1}{2}<\gamma \leq \frac{2 \beta -1}{14 \beta -4} $ &$r_0>\sqrt{\frac{14 \beta  \gamma -2 \beta -4 \gamma +1}{(9 \beta -2) (2 \gamma +1) C}}$& \makecell{$R_0\leq \frac{-2 \sqrt{5}}{\left|R_1\right|} $\\$\mbox{or}$\\$ R_0>\frac{2 \sqrt{5}}{\left|R_1\right|}$}\\
                        \cline{2-7}
                         & \multirow{2}{10em}{\centering \makecell{$k=1$\\ $R(t)=\frac{1}{R_0}-\frac{R_0}{4}\left(\pm t + R_1\right)^2$ \\ \\ $B(r)=r_0
                                        \, \left(\frac{r}{r_0}\right)^{\frac{3 (3 \beta -1) (\gamma -1)}{-\beta  (5 \gamma +7)+\gamma +2}}$}} & \multirow{2}{5em}{\centering$\omega =\frac{3-9 \beta }{9 \beta -2}$} &$0< \beta< \frac{1}{6}$ & $\frac{2 \beta -1}{14 \beta -4}<\gamma <\frac{16 \beta -5}{4 \beta -2}$ & $r_0>2 \sqrt{\frac{14 \beta  \gamma -2 \beta -4 \gamma +1}{(5 \beta  \gamma +7 \beta -\gamma -2) \left(R_0^2 R_1^2+4\right)}} $& \multirow{2}{5em}{\centering \makecell{$R_0\ne0$ \\ $ R_0\ne \frac{\pm 2}{|R_1|}$}}\\
                        \cline{4-6}
                        & & & $\beta <0 $& $\frac{2 \beta -1}{14 \beta -4}<\gamma <\frac{16 \beta -5}{4 \beta -2}$ &$ r_0>2 \sqrt{\frac{14 \beta  \gamma -2 \beta -4 \gamma +1}{(5 \beta  \gamma +7 \beta -\gamma -2) \left(R_0^2 R_1^2+4\right)}} $& \\
                        \hline
        \end{tabular}}
        \caption{A list of some particular solutions with their
                constraints}
        \label{tab1}
\end{table*}
\FloatBarrier

\newpage
\textbf{Data Availability Statement:} No Data associated in the manuscript.
\newcommand{\bibTitle}[1]{``#1''}
\begingroup
\let\itshape\upshape
\bibliographystyle{plain}

\end{document}